\documentclass[conference]{IEEEtran}
\ifCLASSINFOpdf
  \usepackage[pdftex]{graphicx}
\else
  \usepackage[dvips]{graphicx}
\fi

\usepackage{caption}
\usepackage{subcaption}
\usepackage[ruled]{algorithm2e}

\usepackage{url}

\begin{document}
%
\title{CloudFridge: A Testbed for Smart Fridge Interactions}

\author{\IEEEauthorblockN{Thomas Sandholm\IEEEauthorrefmark{1}, Dongman Lee\IEEEauthorrefmark{1}\IEEEauthorrefmark{2}, Bjorn Tegelund\IEEEauthorrefmark{2}, Seonyeong Han\IEEEauthorrefmark{2}, Byoungheon Shin\IEEEauthorrefmark{2}, Byoungoh Kim\IEEEauthorrefmark{2}}
\IEEEauthorblockA{Collaborative Distributed Systems and Networks Lab\\
\IEEEauthorrefmark{1}Graduate School of Culture Technology and \IEEEauthorrefmark{2}Department of Computer Science\\
Korean Advanced Institute of Technology (KAIST)\\
Daejeon, Korea\\
\{sandholm,dlee,tegelund,shan8017,shinbyh,cmossetup\}@kaist.ac.kr
}}


%


\maketitle

\begin{abstract}
We present a testbed for exploring novel smart refrigerator interactions,
and identify three key adoption-limiting interaction shortcomings of state-of-the-art 
smart fridges: lack of 1) user experience focus, 2) low-intrusion object recognition and 2) automatic item position detection. 
Our testbed system addresses these limitations by a combination of sensors, 
software filters, architectural components and a RESTful API to track interaction 
events in real-time, and retrieve current state and historical data to learn patterns 
and recommend user actions. We evaluate the accuracy and overhead of our system 
in a realistic interaction flow. The accuracy was 
measured to 83-88\% and the overhead compared to a representative state-of-the-art barcode
scanner improved by 27\%. We also showcase two applications built on top of our testbed,
one for finding expired items and ingredients of dishes; and one to monitor your health. The
pattern that these applications have in common is that they cast the interaction as
an item-recommendation problem triggered when the user takes something out.
Our testbed could help reveal further user-experience centric 
interaction patterns and new classes of applications for smart fridges that 
inherently, by relying on our testbed primitives, mitigate the issues with existing approaches.
\end{abstract}



%
\IEEEpeerreviewmaketitle

\section{Introduction}
Smart fridges have been designed and built ever since the 90s \cite{kuniavsky2010}, but 
the commercial success and household penetration is negligible compared to 
regular refrigerators. This is surprising given the massive success of technology adoption
in other domestic areas, such as TVs, phones, ovens, and washing machines.
As technological advances make new features available at a lower cost one would expect that
consumers would just adopt these new technologies at little or no cost but with added
value. A number of historical reviews have tried to explain why \cite{harper2003,kuniavsky2010},
but despite the clear advise to focus more on users than technology we see the latest
breed of smart fridges being differentiated from their regular counterparts by nothing
more than a touch screen attached to the door with stock tablet features.
This epic failure of meeting user expectations is also very apparent in blogs and reviews
of smart fridges after they are introduced at shows like the Las Vegas Consumer Electronics Show~\footnote{\scriptsize{e.g. \url{http://www.theguardian.com/lifeandstyle/2012/jan/11/homes-fooddrinks}, and \\ \url{http://www.theconnectivist.com/2013/04/how-smart-do-we-need-our-homes-to-be/}}}.

Although it is easy to dismiss the commercial failure as the price point being wrong, the constant 
stream of failures across companies and markets over the last 20 years tells a different story.
Instead, if we assume the issue is really a lack of focus on the users, and improving the user experience
is the solution, what can we do to remedy the situation?

First we need to understand how refrigerators are used. A number of studies help us here. E.g. Paker and
Stedman~\cite{parker1992} found that people interact with the fridge 42 times a day by opening and
closing the door and Cusack et. al.~\cite{cusack2012} found that 70\% of the users in their study
listed due item removal notifications as the most desirable feature of a smart fridge.

Now, in order to test alternative interaction designs {\it in situ} we propose
\emph{CloudFridge}, a testbed that can be easily deployed in regular fridges to
give them smart capabilities. The testbed can be used both by sensors and applications
to coordinate and track interesting interaction-related events.

Here we present the design, implementation, deployment and evaluation of CloudFridge to support
interactions that 1) are user-experience centric 2) provide item recognition that is 
non-intrusive to the simple, short and frequent user interactions occurring today, 
and 3) track removals of items by automatically detecting the position in the fridge.

The design follows the principle of combining state-of-the-art, off-the-shelf hardware and 
software building blocks and exposing them using a RESTful Web service interface that can be used
both for notifications, as well as current and past state queries by applications.
Key components of the testbed architecture exposed by the Web service include Phidget InterfaceKit Sensor Boards with
IR proximity and force sensors and LED actuators; an Odroid-X2 Single Board Computer (SBC) with a HD webcam and Internet connectivity;
and a cluster of Google Search By Image video frame object recognizers
deployed in a dynamically scaling Cloud infrastructure.



Our contribution in this paper comprises: 1) a testbed system to help prototype realistic, interactive smart fridge applications, including tracking usage, getting current and past state, and real-time notifications of events; and 
2) two use cases of applications built on top of our testbed to showcase novel interactions.

The rest of this paper is organized as follows. In Section~\ref{sec:related} we discuss previous efforts to design smart fridge
systems. We then formulate the requirements of our testbed in Section~\ref{sec:requirements}. Section~\ref{sec:design} presents an 
overview of our testbed system design.
Section~\ref{sec:evaluation} evaluates the core primitives of our testbed, and Section~\ref{sec:applications} exemplifies
how  user experience centric interaction design can be facilitated by our testbed by presenting two smart fridge applications
based on our system. Finally, in Section~\ref{sec:conclusion} we conclude. 

\section{Related Work}\label{sec:related}
Smart fridges were introduced as part of the smart home concept in the late 1990s. 
Today a smart fridge is expected to
self-monitor food items and expose a user-friendly interface for general users~\cite{gangadhar2011,luo2009}. 

Rouillard~\cite{rouillard2012} proposed a pervasive fridge system to avoid food waist. 
Their system exploits smart phone capabilities for input of food items to lower the cost, including
barcode scanning and automatic speech recognition. 
The main drawback of this type of interface is that it is intrusive and requires significant 
user intervention. They also propose a proactive deadline notification system using SMS and 
instant messages. However, unless the user is about to use 
the item or is in a grocery store the notification may easily be forgotten and ignored as well, i.e. 
the alerts are not aware of or sensitive to the context or the current situation of the user.  

Gu et al.~\cite{gu2009} implemented a smart fridge system to help people 
adopt a well-balanced eating habit. The food items stored in the refrigerator are scanned 
by an RFID reader and the information is sent to a remote server to create a shopping list. 
This method assumes that food items have an RFID tag, which is rarely the case. 

Luo et al.~\cite{luo2009} focus on  
healthy nutritional habits and provide a system for automatic generation of shopping lists and 
recipe recommendations based on food stored in the refrigerator. 
Food items are recognized via a touch screen interface, which involves user intervention.  

Murata et al.~\cite{murata2012} proposed a method to detect wasteful usage of a fridge by observing the 
door opening and closing. The refrigerator is equipped with a distance sensor to 
detect a person approaching the door, a magnetic contact switch to detect door opening and closing events, 
and infrared distance sensors to detect items being put in or taken out. However, the exact position
of items being inserted or removed is not recognized, which is the key to our approach.

Bonanni et al.~\cite{bonanni2005} propose a series of Augmented Reality projected cues to 
guide the user through various tasks in the kitchen. They project the content
inside the fridge on the outside of the door to avoid having to open it to look
for an item. They also project light arrows that can point to the position of an item that
you are looking for. A very similar 
system was proposed by Ju et al.~\cite{ju2001} where the purpose of the AR projected guides
was to teach users how to cook, e.g. by showing helpful video snippets directly on the
kitchen counter where the food is being prepared.
This work on user interface design is complimentary to our work as we focus
more on the core interaction capabilities of the fridge and tracking activities to find patterns
that can in turn be used for recommendations, or to guide users through tasks.

\section{Testbed Requirements}\label{sec:requirements}
Based on the preceding analysis of gaps in state-of-the-art approaches
to smart fridge designs we now formulate a set of requirements
for a testbed system that addresses these issues.

\begin{itemize}
\item {\bf R1: User experience centric.} Previous approaches
have been technology centric, and even though the technology
issues have been solved, the user experience issues have
surprisingly remained similiar over the years in both commercial and research
lab work. Hence, our main claim in this paper is
that a testbed geared towards evaluating realistic user experiences
could be the catalyst for innovations in the field.
\item {\bf R2: Low-intrusion object recognition.}
Technology supported features and day-to-day interactions with real-world
articfacts have been poorly integrated leading to manual high-intrusion
user intervention to fill the gap. The concrete example of that is that
today's smart fridges cannot automatically recognize what item you put in.
Solving that problem is key to enable any kind of smart interactions.
\item {\bf R3: Automatic item position detection.}
On a related note, if we detect the exact position of an item put into the fridge 
we can easily correlate it with the object recognized during input. This 
avoids having to recognize the object twice.
Furthermore, we open the door for applications that help users find items
in the fridge, and track the consumption time as well as duration stored of items to learn user and item
preferences, and correlations respectively.
\end{itemize}

Next we describe how our testbed was designed to meet these requirements.

\section{Design}\label{sec:design}
The CloudFridge testbed provides a Web service interface both for end-user applications running
on consumer devices and for fridge sensors that are deployed in each fridge.
The Web service runs on a server in the Cloud and is thus generally accessible
from any Internet connected device.

\begin{figure}[htp]
\centering
\includegraphics[scale=0.35]{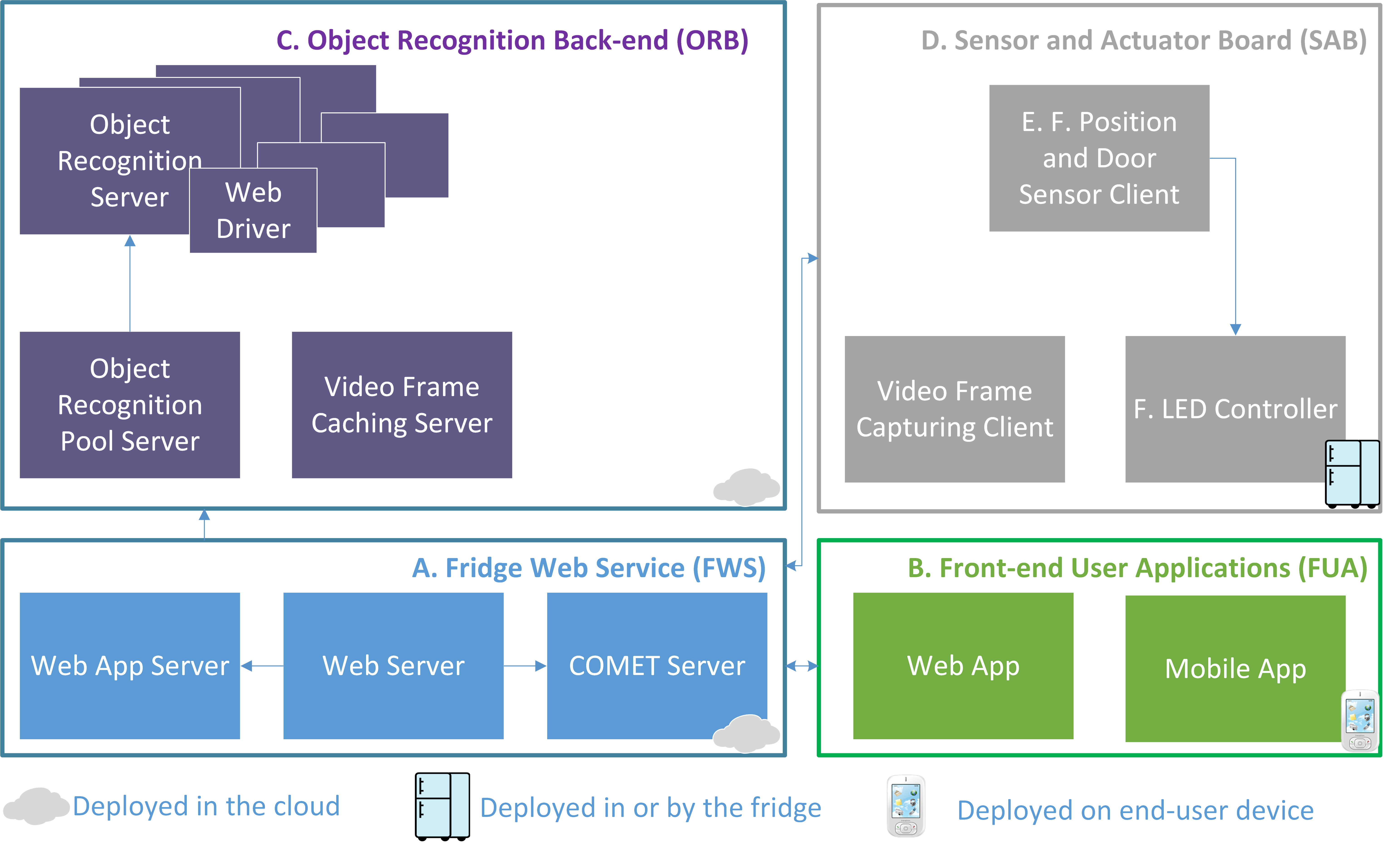}
\caption{CloudFridge architecture.}\label{fig:architecture}
\end{figure}

Figure~\ref{fig:architecture} shows an overview of the CloudFridge architecture
and the interactions between the system components. 
Component interactions always go through the Fridge Web service (FWS)
both for synchronous requests and asynchronous events. For example
some sensors may listen on events from other sensors to trigger or stop
sampling activities.

The Object Recognition Back-end (ORB) would typically run in a compute cluster as it is 
the most resource intensive
component. The Sensor and Actuator Board (SAB) is deployed in each fridge in the system. 
All sensors belonging to the same fridge 
communicate on the same publish-subscribe event channel. Each fridge thus operates
completely independently of other fridges. Fridge exchanges have to be done on
an application level, Fridge User Applications (FUA), if allowed.  This shared-nothing architecture at the core 
allows for efficient load balancing and partitioning. 

We now describe the design
of each CloudFridge component in some more detail.
\subsection{Fridge Web Service (FWS)}
The FWS comprises a Web App Server (Django~\footnote{https://www.djangoproject.com/}) to handle queries about fridge state 
and item meta data~\footnote{including recognizing items in images}, and a COMET 
Server~\footnote{https://github.com/jedisct1/Simple-Comet-Server}
to send out push notifications about important events (e.g. item put into fridge).
The Web Server component is a standard HTTP Server (Apache) which exposes a single uniform 
JSON(P)/REST API and serves as a proxy to to all FWS components. The FWS 
components are supported by a local MySQL database which thus completes
the LAMP~\footnote{http://en.wikipedia.org/wiki/LAMP\_(software\_bundle)} stack architecture.   

\subsection{Front-end User Applications (FUA)}
Applications that want to talk to our Fridge service need to support
HTTP POST, GET and JSON marshalling. 
For WebSocket style notifications we use a 
long-polling client that interacts with the FWS COMET component via the REST API.
The advantage is that it works both in Web and native code without any special purpose 
client libraries. We have written clients in Javascript (JQuery), Python, and Java (Android) as part
of the system infrastructure. All fridges are identified with a unique {\it Fridge ID}. That ID
is all that the clients need to get access to 
the current content of the fridge, the history of actions, and events pertaining to 
input and output actions and alerts as they happen. Hence, during social sharing, this ID
is all that needs to be shared.  

\subsection{Object Recognition Backend (ORB)}
Object recognition is done on a frame-by-frame basis where as many as 5 frames per second may be
sent to the server for recognition. Given that the object recognizer that we use, Google 
Search By Image, has a response time of 2-5 seconds for a single image it meant that we had
to parallelize the recognition step, as well as perform it in the background to avoid 
affecting the end-user experience. A pool server hands out leases to 
handles to back-end recognizers. When the app server gets the result back it releases the lease
for another request. All this communication is done using Pyro as all components are implemented
in Python, but they run in different processes and potentially also on different physical hosts.
On our current testbed we make use of 9 recognition servers on a single machine, 
but servers can be added and removed from the pool at any time without having to restart the system.

We decided to use the Google Search By Image feature since it does not require any training,
and it is resonably accurate and fast (see our evaluation section below).
There are however two drawbacks with this choice, which is why we have left the architecture open to
plug in other object recognizers either to complement the current recognizer or to replace it.
The first drawback is that the service only recognizes texts and logos accurately. Pure image searches
typically only returns images with the same color palette, or identical images if the image
is already online, neither of which is useful to us. However, the generality of the solution
was appealing to us and it does go one step beyond barcode scaning since the item does not need
to have a barcode first of all and second you do not need to find the barcode and turn it
towards a scanner. Therefore this solution reduces the overhead which is a key concern for us. 

The second more tedious drawback is that there is currently no API to Google Search By Image.~\footnote{
Some hackers have tried to reverse-engineer wiredumps and discovered that it is a ProtocolBuffer
protocol. These solutions were too unreliable to us, as they break whenever Google modifies their backend.}
To solve this problem we utilize a Web driver that performs the search inside a Web browser. 
We maintain the session
to the Web browser in memory (in the Pyro server) so that we only need to start a browser 
once during startup. We make use of the Selenium Web driver~\footnote{\url{http://docs.seleniumhq.org/}} from a python client from the Pyro server. 
This setup allows us to get virtually the same remote call response time as the interactive response
time from a browser. Currently we extract
the ''best search key`` value of the result, which gives the best key words to use to find similar images.
This works well because it does not have many false positives. The one drawback is that the same
item can return slighly different search keys. This problem is easy to solve with a regexp filter. 
We optionally allow only items that match a regexp filter to be returned in which case you can 
almost completely eliminate false positives. The advantage of not relying on the regexp filter is
that new unknown items can be recognized without any pre-configuration. 
One could imagine a semi-automated process where the end-user receives the raw key words if the
item does not match a regexp, and can select a suggested canonical item description and some 
suggested terms to match the description with. 
Many of the use cases and even some parts of the core infratructure compoments work better
if canonical representations of the same item can be assumed.

As a final piece in the puzzle we provide a video frame image caching service, also
remotely accessible through a Pyro interface. The main purpose of this service is to allow the
Web driver to simply feed in URLs to the video frames into Google Search By Image,
which simplifies scripting and avoids writing images to disk. This is useful in our
scenario because most of the video frames may not contain
any interseting objects or there may be many duplicate frames of the same object. Google will then fetch 
the frames during the recognition
phase, which is the reason why we need to generate a short-term public URL to the frame. 
If an item is recognized in a frame we create a permanent image file on disk
that is accessed by a separate URL to serve our end-user applications.
An additional advantage of this design is that in the case of a pipeline of multiple
independent object recognition services and algorithms they can all access the
temporary URL and the images could still be kept in the cache in one place.

\subsection{Sensor and Actuator Board (SAB)}
There are three types of sensors (visual, proximity, force) and 
one actuator (LED) that our system interacts with.
The visual sensor is simply a HD web camera. We extract video frames
when we detect activity (a user opens the door) until the activity stops
(a user closes the door). The frames are all uploaded in parallel to the
fridge service for recognition. If an item is recognized it is reported
with an activity ID as a new item to add to the fridge. 
The activity ID is reset every time the door opens to
distinguish between duplicate recognitions and new objects. There is
also a timeout so if the time difference between two successful object
recognitions is long enough (e.g. 10 seconds) the objects are also
considered to be different. This is important since you may put in
two instances of the same item type, e.g. two coke bottles.   

The proximity sensor is used to detect where an item was placed
and from where an item was removed. This process runs independently
and in parallel to the object recognition process. If the object 
is recognized before the position, a pending item is added to our data
base without a position, and then when the position is detected it will be associated
with the last known pending item. Conversely if the position is detected first
a placeholder item will be added in that position in the fridge. When an object
is recognized it will replace the placeholder in the same position.
Relying on the door opening and closing to make sure that the postion
recognition is matched to the correct object recognized is unreliable
because sometimes the door may close before the object recognizer is
done analyizing all frames, which is why we allow these events
to happen asynchronously. 

The force sensor detects when a door closes or opens which is
used to know when video frames should be captured and analyzed. This
is important because it is a heavy operation both for the sensor board,
the network and the server, so continous recognition is not feasible.
Different applications may also want to react to this event to, for instance,
serve the user with recommendations when the door opens.
Finally, the LED actuator is used to light up positions in the fridge to
direct the users attention there. Currently we use red and green lights.
Red lights mean alerts, e.g. an item expired. Green lights mean recommendation,
e.g. the application recommends that the item could be consumed for various reasons.
Exactly what semantics is associated with the lights is up to the applications.

\subsection{Position and Action Detection}\label{sec:implementation}
The position and action (add, remove, or none) detection
component is central to our approach and design as it
is important to obtain maximum accuracy in this piece for
our system to work properly as a whole, which is why we describe
this component in more detail below. 
After testing a large variety of sensors to detect position including
force, vibration, and sonar sensors, we chose IR proximity sensors because they
were most robust to the fridge environment. For instance, they were not affected by humidity. 
One issue with IR sensors is that they record different values whether
the object detected is metallic (reflective) or not. Non reflective objects
result in higher values than no items and reflective items yield lower values.
We will describe next how our software filter addresses this issue and others.


The first filtering of the raw sensor values 
was to not simply fire detection events if a threshold was exceeded,
or in this case if the proximity fell below a value, but to average
multiple readings. The reason is that it is quite easy to fire an
event from the neighboring sensor because your hand is in the way
or you do not put the item down in a straight line. We compute
a sliding-window, moving-average mean of 5 readings for this reason. We typically
get about 1 reading per second with the sensors we use. This means
that you need to obstruct a sensor by mistake for more than 5 seconds
for it to potentially fire an event. The second filter 
is to more reliably determine whether someone took an item out or put one in.
If you compute the averages, they may look the same over five seconds
for both adding and removing an item. The first step in solving this
problem is to simplify matters by only considering events that happen within
the same activity period. Recall that activity periods are framed by
opening and closing the door. Now within the activity period we compute
and track three moving average (5s interval means) values, the minimal
value, the maximum value and the last value computed before the door closes.
The minimal value is very stable and a good indicator
for adding a reflective (metallic) item, and the maximum value is a good indicator
for adding a non-reflective item.
These values essentially determine whether an item is currently
in a position. 
This is needed both to detect additions of items but also
to avoid falsely detected removals. The last read interval mean is on the
other hand a stable indicator of a removal or if an item is not in a 
position anymore (and to avoid falsely detected insertions).
We signal a removal if the last value is higher than the threshold for
reflective items and lower than the threshold for non-reflective items (see Algorithm~\ref{alg:actposalg}).


To deploy the sensors in a new fridge we calibrate the 
minimum (reflective), $it_{min}$, and maximum (non-reflerctive), $it_{max}$, average thresholds to
fire input events. We also configure last average thresholds for reflective, $ot_{min}$,
versus non-reflective, $ot_{max}$, items to signal output events.

This solution is expandable to situations
where you want to detect positions of items stacked in the width, depth, as well
as hight dimensions. If a series of proximity sensors are installed along the 
walls vertically as well as horizontally you could create a three-dimensional
grid that could potentially recognize more sophisticated positions. 

\begin{figure*}
\centering
\begin{subfigure}{.3\textwidth}
 \centering
 \includegraphics[scale=0.42]{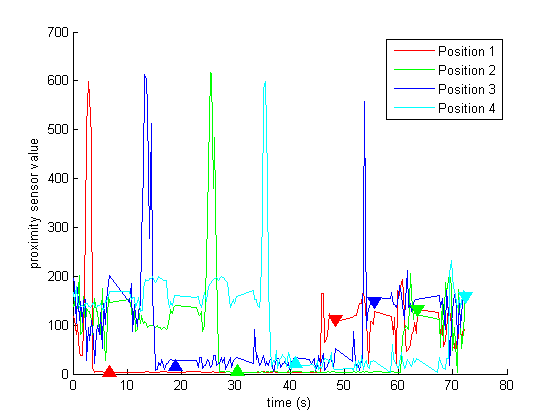}
 \caption{Soda cans (reflective).}
 \label{fig:sodapos}
\end{subfigure}
\begin{subfigure}{.3\textwidth}
 \centering
 \includegraphics[scale=0.42]{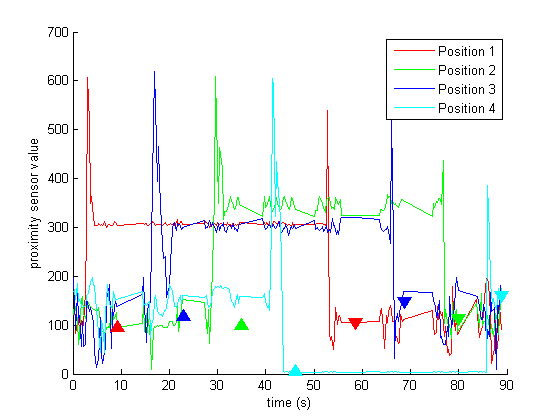}
 \caption{Milk packages (non-reflective).}
 \label{fig:milkpos}
\end{subfigure}
\begin{subfigure}{.3\textwidth}
 \centering
 \includegraphics[scale=0.42]{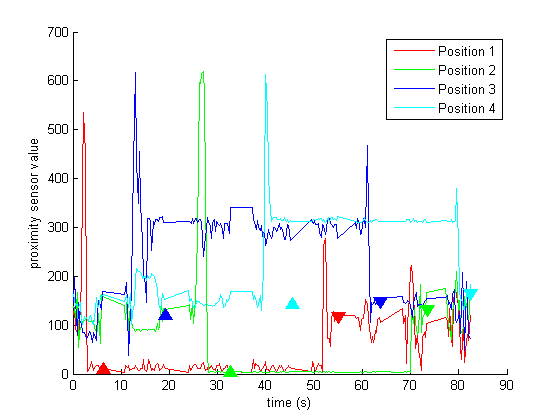}
 \caption{Soda cans in position 1,2 and milk packages in 3,4.}
 \label{fig:mixpos}
\end{subfigure}
\caption{Sensor values and our system's decisions that items are being added (upward-pointing triangles) and
removed (downward-poinintg triangles) for reflective and non-reflective items.}
\label{fig:positionsensor}
\end{figure*}

Figure~\ref{fig:positionsensor} shows measured proximity sensor values and our system's decisions in
three example runs where items were put in positions 1,3,2,4 and then removed from positions 1,3,2,4.
The first run used reflective items only (soda cans), the second run used non-reflective items only
(milk packages), and the third run used a mix (first and third item soda cans and second and forth items
milk packages). We can see that the lower thresholds are used for the soda cans and the higher thresholds
for the milk bottles. There is an exception though that we cannot fully explain. The last milk bottle in
the all milk run does reflect. However, this is not an issue in our case since it is then just treated
like a reflective item.
We note that the add (upward-pointing triangle) and remove decision points (downward-pointing triangle) are clearly separated, 
to avoid false positives and false negatives. We also
note that the add and remove thresholds are very similar across the positions and sensors
which shows that it is easy to calibrate and configure as the same thresholds can be used for all positions.


\begin{algorithm}
\SetAlgoLined
\For{all sensor positions}{
  \If(){$(minval < it_{min}$ {\bf or} $maxval > it_{max})$ {\bf and} no item in position}{
   add item in position
  }{
  \ElseIf{$ot_{min} < lastval < ot_{max}$  {\bf and} item in position}{
   remove item from position
  }
  }
  
}
\caption{Action and position determination. Min thresholds take effect for reflective items and max thresholds for non-reflective items. Here $ot$ is output threshold and $it$ is input threshold.}\label{alg:actposalg}
\end{algorithm}

\subsection{Hardware Architecture and Deployment}\label{sec:hwdesign}
\begin{figure}[htp]
\centering
\includegraphics[scale=0.3]{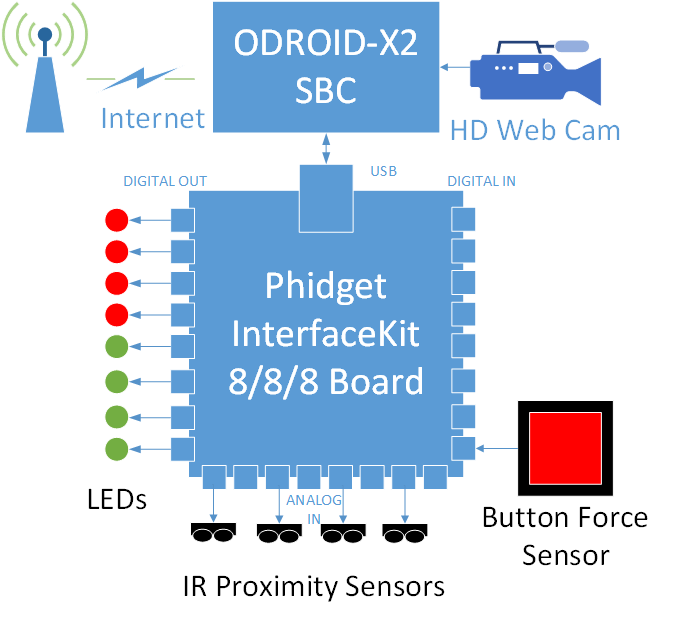}
\caption{CloudFridge hardware component architecture.}\label{fig:hwarchitecture}
\end{figure}

Figure~\ref{fig:hwarchitecture} shows a schematic of the
sensor board and single board computer (SBC) design. The sensor board and
all the sensors are inside the fridge, wheras the SBC is outside. On the
SBC the wireless or wired connection is outside whereas the web camera
is inside. 
There may be many sensor boards (Phidget Interface Kit)
connected to a single SBC (Odroid-X2). Each sensor board has 8 LEDs and 4 IR proximity
sensor connected to it and can thus serve 4 positions. One of the sensor boards
also serves the force sensor to detect door activities in one of the digital input slot.
The IR proximity sensors are connected to the analog inputs, and the 
LEDs are connected to the digital output ports.

\begin{figure}[htp]
\centering
\includegraphics[scale=0.3]{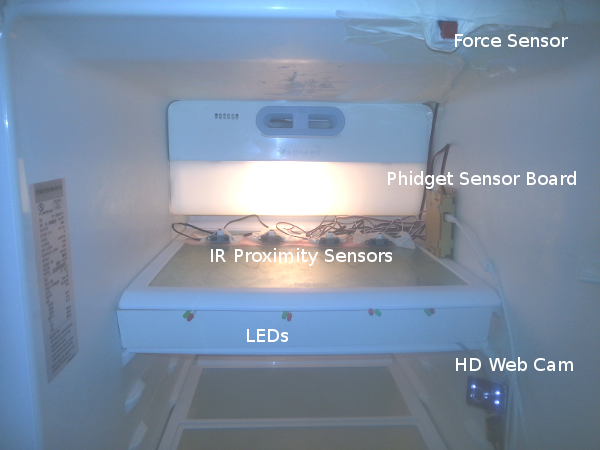}
\caption{CloudFridge hardware deployment inside of fridge.}\label{fig:fridge}
\end{figure}

Figure~\ref{fig:fridge} shows our actual deployment inside
our test fridge. We only need to pull a USB cable through the fridge door, which is
very straightforward given the insulating material attached to the door edges.
The operational temperature of the sensors and the Phidget board is 0-70C, so it 
works well in the fridge. However, for a freezer deployment at least the sensor
board needs to be outside, although some special purpose sensors may work in colder
temperatures too. Note that this design could easily be generalized and used in
any other item storage space too. For instance on a book shelf. Books are recognized very
easily by our object recognizer as well, as they almost always have plenty of text on the cover.

\section{Core Interaction Flow Evaluation}\label{sec:evaluation}
In this section we evaluate mainly how well the testbed meets requirements {\bf R2} and
{\bf R3} from Section~\ref{sec:requirements}, i.e. low-intrusion object recognition
and automatic position detection. Requirement {\bf R1} of user experience centric design of realistic
interactions is also adhered to by using user centric metrics such as accuracy and overhead 
to evaluate these features in a core~\footnote{a flow that is a subset of all application interactions} interaction flow. 
However, this requirement is the main focus of the use case section (Section~\ref{sec:applications}).

We are primarily interested in measuring the accuracy of
item, action, and position recognition.
Having an automatically generated
trace of timestamped item, action, position events
will provide many opportunities for innovation in terms of
new ways to interact with the fridge, as we will exemplify in Section~\ref{sec:applications}. 
Now the question is: is it accurate enough to provide
any added value over a regular fridge without too much
intrusion and overhead? To this end we study the
accuracy at different levels of overhead.

\subsection{Metrics}
\begin{figure}[htp]
\centering
\includegraphics[scale=0.48]{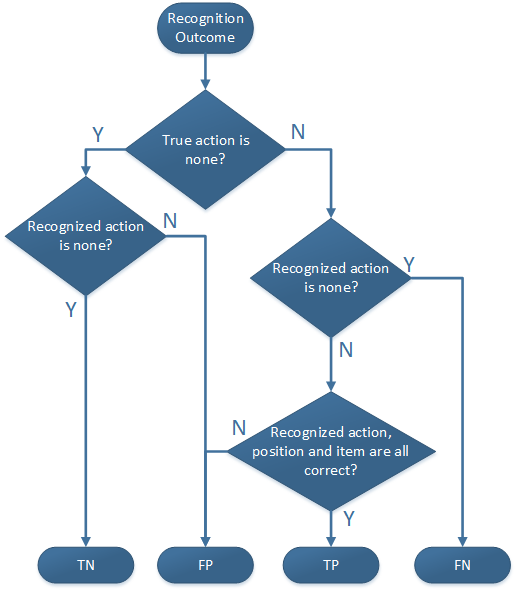}
\caption{Mapping of experiment outcomes to truth values used in precision and accuracy metrics.}\label{fig:decisiontree}
\end{figure}

For overhead we simply measured the time between opening
and closing the door for various actions compared to
using the fridge normally.
To determine how well we recognize item actions, we first define precision
as:
\begin{equation}
   P = \frac{TP}{TP+FP}
\end{equation}
positive is defined as any add or remove action, and negative is defined
as any none action, i.e. a user opening and closing the door without
taking anything out or putting anything in. Hence the precision, $P$,
ignores how well we predict the none actions and only looks at how well we detect true additions
and removals. ${TP}$, and ${FP}$ here refer to True Positive and False Positive,
respectively. To see how sensitive our system is to none actions we define
accuracy as: 
\begin{equation}
   A = \frac{TP+TN}{TP+FP+TN+FN}
\end{equation}
here ${TN}$, and ${FN}$ refer to True Negative and False Negative respectively.
To simplify the visual presentation in graphs we also define Precision Error as:
\begin{equation}
   PE = 1-P
\end{equation}
and Accuracy Error as:
\begin{equation}
   AE = 1-A
\end{equation}
We note that both $PE$ and $AE$ are still in the range $[0,1]$.
For a complete definition of
how we map recognition results to truth values see the decision tree in
Figure~\ref{fig:decisiontree}.

\subsection{Experiment Setup}
\begin{figure}[htp]
\centering
\includegraphics[scale=0.35]{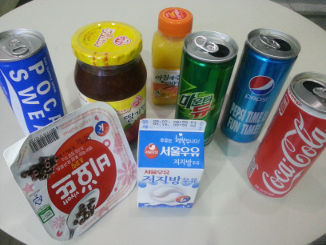}
\caption{Items used in the experiment.}\label{fig:items}
\end{figure}

We generate a random script of add, remove and none actions in a sequence. The add
actions also randomize which item to add from a list of our eight experiment items 
(see Figure~\ref{fig:items}) and which position to put the item in, out of
the four positions available in our single sensor board deployment. 
The add and remove actions are state-aware, i.e. an item is always put
in an available position and removed from its last position.
Given these constraints the script is completely random. The script comprises
50 steps. Recall that a user interacts with the fridge about 40-50 times a day according to~\cite{parker1992},
which made this number appropriate for our experiment.
Now for each step we keep track of the time the door was open and the event our system
recognizes, and compare it to the ground truth event from the script.
To get a baseline for the overhead we also ran through the script without using our system except
for the door sensor and measured the time it took between opening and closing the door for each
step in the script. Furthermore we generated a second random script as a baseline for the precision
and accuracy metrics.
\subsubsection{Statistical Evaluation}
Given that we want to compute overheads at different levels of precision and accuracy we
could run the experiment with 50 actions many times and compute the average and variance of
overhead across experiment runs. However, to save time we employ a statistical
bootstrap~\cite{efron1982} inspired sampling procedure that allows us to get estimates
of overhead versus precision and accuracy from one single experiment run.
It works as follows. We take 100 random subsamples (without replacement) with 10 steps each.
For each random step we compared the baseline duration (between door opening and closing)
to the duration of the real experiment run to get the overhead. Then we computed the
precision and accuracy values across the 10 steps. Our final evaluation measures are then various average
and variance values across these 100 subsamples.
\subsection{Precision Results}
We run two flavors of our experiment. One using only the four reflective items (soda cans) and one
using all eight items (including the soda cans and non-reflective items). One reason for
this setup is that our position sensor runs a more complex algorithm if
items are not reflective (see Section~\ref{sec:implementation}). It also allows us to see how
our system scales in terms of recognizing a larger pool of more varied candidate items.

To visualize the precision and accuracy at different levels of overhead we computed the
average and variance of all subsamples with an overhead less than $x$ seconds and let
$x$ vary\footnote{there were too few samples to start $x$ at 1} from $2$ to $10$.

\begin{figure}[htp]
\centering
\null
\begin{subfigure}{.2\textwidth}
 \includegraphics[scale=0.26]{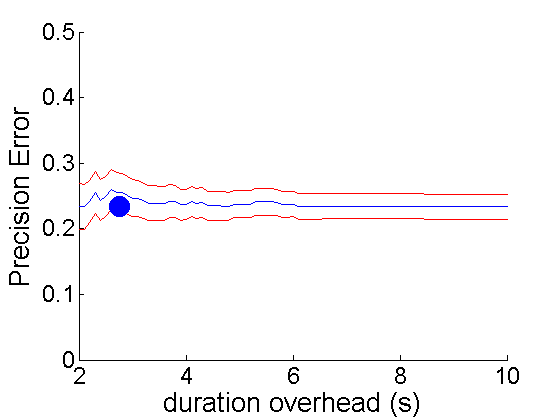}
 \caption{Reflective items.}\label{fig:sodaprecision}
\end{subfigure}
\hfill
\begin{subfigure}{.2\textwidth}
 \includegraphics[scale=0.26]{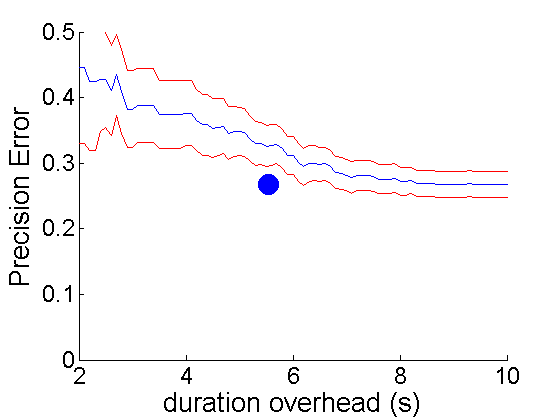}
 \caption{Mixed items.}\label{fig:mixprecision}
\end{subfigure}
\hfill\null
 \caption{Precision Error versus Overhead for recognizing the correct item, position as well as action.
The blue circle pinpoints the average Precision Error at the average duration. The red lines
show the standard error range.}\label{fig:itemposaction}
\end{figure}

The results for precision error can be seen in Figure~\ref{fig:itemposaction}.

The average precision for the 4-item-reflective versus the 8-item-mixed experiment
are very close (0.77 versus 0.73). This difference can be explained by the position
accuracy degrading in the mixed experiment. Item recognition accuracy was stable. 
In the 4-item experiment all errors are due to item recognition failure, and position and
action precision are both perfect (100\%). Looking at the overhead we see that the mixed experiment
shows a clear trend of precision improving with overhead, but at the average level
of overhead the precision is already close to saturating. Similarly for the reflective
experiment the precision saturates close to the average duration, but the effect of overhead
is not as pronounced.  Not surprisingly the overhead is larger for the mixed experiment both in
absolute terms and proportionally to the baseline duration (79\% versus 65\%).

\subsection{Sensitivity to False Negatives}
Next we study the accuracy metric to determine how sensitive our system is to false negatives~\footnote{we predict a none action when in fact something was added or taken out}.
\begin{figure}[htp]
\centering
\null
\begin{subfigure}{.2\textwidth}
 \includegraphics[scale=0.26]{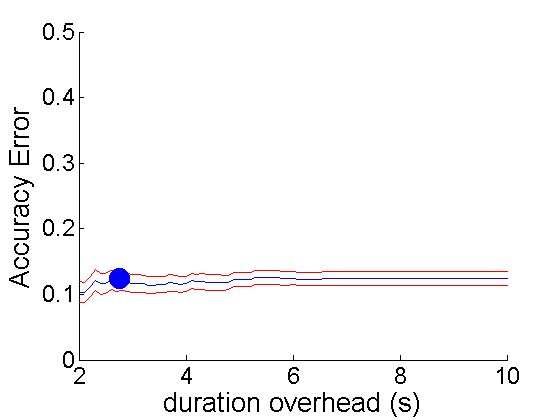}
 \caption{Reflective items.}\label{fig:sodarecall}
\end{subfigure}
\hfill
\begin{subfigure}{.2\textwidth}
 \includegraphics[scale=0.26]{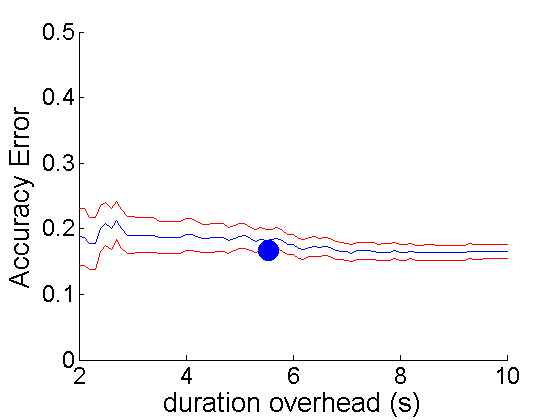}
 \caption{Mixed items.}\label{fig:mixrecall}
\end{subfigure}
\hfill\null
 \caption{Accuracy Error versus Overhead for recognizing the correct item, position as well as action.
The blue circle pinpoints the average Accuracy Error at the average duration. The red lines
show the standard error range.}\label{fig:itemposactionrecall}
\end{figure}
Figure~\ref{fig:itemposactionrecall} shows Accuracy Error for the two flavors of our experiment.
For both the reflective and mixed experiments the accuracy is significantly lower than
the precision, which indicates that the system is not very sensitive to false negatives.
In the reflective experiment there was not a single incorrect add or remove event resulting from
a none action. We also note that accuracy is not sensitive to overhead in any of the experiments.
The absolute average accuracy values for our experiments are also encouragingly 
high: 0.83 (mix) versus 0.88 (reflective).



\subsection{Improvement over Random Baseline Results}
As a sanity check we compared the average precision and accuracy values for a random baseline that produces
a valid series of steps but completely randomly. The precision was 0\% for both experiments and the 
accuracy 0.16\% versus 0.33\% for the reflective versus the mixed experiment. The reason
why the accuracy is relatively high is that we have as high probability for a none action as an add and remove
action in the experiment, and because a none event has an implied position (none) and item (none).
So a biased predictor that just predicts dummies all the time will have similar values. Regardless,
our recognizer has precision and accuracy values that are substantially higher. 


\subsection{Barcode Scanning Comparison Results}
Given that many existing solutions use barcode scanners to recognize items and that the items used
by this experiment have barcodes, we also implemented an Android barcode scanning application that 
makes intent callouts to a barcode scanning app called ZXing~\footnote{https://code.google.com/p/zxing/} to search 
images for barcodes. The only thing the user has to do is to press a button to start the scan and then
observe the fit in a standard scanner camera view of the smartphone. Once a fit is found by ZXing we will
get a notification or intent result back into our app with the recognized text. We then automatically call
our Fridge Web service to register the recognized item with the fridge. Hence, we believe that this is a
reasonable approximation of the user experience and overhead of state-of-the-art barcode scanning fridges.
Running our experiment with this setup, precision and accuracy were both perfect, but 
barcode-based recognition showed
an average overhead of $4.1$ seconds compared to the baseline~\footnote{we ran the reflective (soda can) experiment with the barcode scanner since barcode scanning does not affect the position accuracy, and the position accuracy is perfect with soda cans.}. This is about $1.4$ seconds higher than our
image recognition method, or a 27\% increase in interaction time. This difference is statistically significant in a t-test (see Table~\ref{T:experiment}).  
The additional overhead can be explained by the extra user intervention involved in starting a scan and fitting the 
barcode of the item inside the area where the scanner can recognize it. One could argue that
barcode scanning could be automated by sensors just like our image recognition sensor, but in our
testing we found that the more general image recognizer was much less sensitive to imperfect video frame captures.
Supporting these imperfections is important to reduce the intrusiveness of the solution. 
Another drawback of barcode scanning is that the barcodes are product and store specific, so they in most cases
do not provide any human readable hints if the product is not already known in our system.
Furthermore, not all products have barcodes or may have them in positions that are hard to find, although this
was not a factor in our experiment since we only used soda cans with the barcode placed at the same place.

\subsection{Results Summary}


\begin{table}
\centering
\caption{Summary statistics for our recognizer in the two experiments. 
P-values are of null-hypothesis (NH) in two sample t-test.}\label{T:experiment}
\begin{tabular}{|l|c|c|} \hline
{\bf statistic} & {\bf soda } & {\bf mix } \\ \hline
Mean precision & 0.76 & 0.73\\
Mean accuracy & 0.88 & 0.83\\
Correct item ratio for add action & 0.77 & 0.82 \\
P-value of $NH_p$ & $3.6\times10^{-97}$ & $2.2\times10^{-93}$  \\
P-value of $NH_a$ & $6.6\times10^{-111}$ & $6.2\times10^{-74}$  \\
Overhead compared to baseline& $0.65$ & $0.79$ \\
Add action overhead & $2.6$ & $1.9$ \\
Remove and Dummy action overhead& $-.19$ & $-0.01$\\
P-value of $NH_b$& $4.4\times10^{-8}$  & - \\
Overhead compared to barcode scanning& $-27\%$ & - \\
\hline
\end{tabular}
\end{table}

We summarize our experimental results in Table~\ref{T:experiment}. 
$NH_p$ is the null hypothesis that our system does not have a higher precision than the random
recognizer, $NH_a$ is the null hypothesis that our system does not have a higher accuracy than
the random recognizer, and $NH_b$ is the null hypothesis that our system does not have a lower
overhead than barcode scanning.

The drop in precision and accuracy in the mix experiment can be attributed to our position
recognizer performing worse with non-reflective items. The increase in overhead in the mix experiment
resulted in a higher add item recognition ratio. In both experiments, all the overhead is
accumulated only in the add actions, while scanning the item. We note, however, importantly that we 
decrease the overhead compared to the barcode scanning approach with as much as 27\%, which is
significant considering that this is an action that happens about 25 times a day assuming half of the actions
are additions of new items.

\section{Use Case Discussion}\label{sec:applications}
The main purpose of this section is to exemplify user experience centric design of interactions
as stated in requirement {\bf R1} (Section~\ref{sec:requirements}). We also provide a preliminary
analysis of common interaction patterns to propose a new class of interactions with smart fridges.
Next we present two end-user applications that we have implemented on top of the CloudFridge testbed system.

\subsection{cfTakeOut Application}
\begin{figure}[htp]
\centering
\includegraphics[scale=0.33]{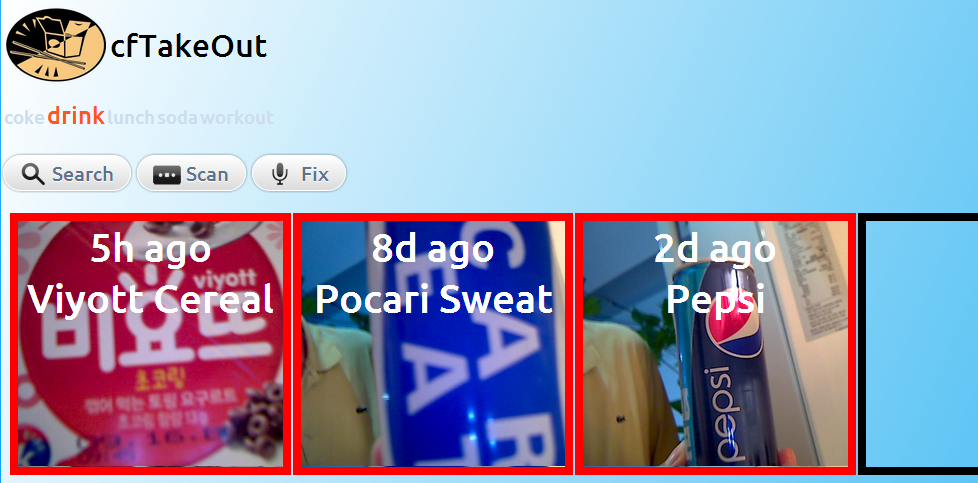}
\caption{Screenshot of cfTakeOut application.}\label{fig:takeout}
\end{figure}

The {\it CloudFridge Take Out} (cfTakeOut) application makes it easier for 
you to find what you want to (or should) take out of the fridge.
You can tag items with type (e.g. drink), occasion (e.g. lunch),
or dish (e.g. pancakes). These tags and the item names themselves 
can be used to search for items by voice when you open the fridge door.
Furthermore, the application keeps track of the average time
you leave an item in the fridge and recommends that you take out
items that have passed this time with a certain margin. It 
also tracks the time-of-day you normally take out certain items
and recommends that you take them out when you open the door during those
times. The recommendations, as well as item positions and names are visible in a virtual fridge in the app
UI (see Figure~\ref{fig:takeout}) but they are also indicated by LEDs in the fridge (red for item expiration
and green for recommendations), and may be spoken out by a Text-to-Speech engine. 
Hence, the user can interact with the
fridge without any direct visual or touch interactions with the device
running the application, which showcases our design of low-intrusion
interactions. Door opening and closing events are also leveraged to
trigger Speech-to-Text recognition to enhance the item recognition~\footnote{when door closes without item being recognized correctly} or to
do voice search~\footnote{to find items in the fridge after you open the door}. 
The app is written using HTML 5 and runs in both mobile and desktop browsers. 
When running on an Android phone this application also allows barcode scanning to be used
to improve recognition accuracy at the cost of increased overhead.

We are planning on extending this app to track correlations between items consumed together to
recommend which item to take out next, and we are also planning on adding social features to
share your fridge content with friends or compare tastes to be able to get 
collaborative-filtering-based and social recommendations.

\subsection{cfHealth Application}
\begin{figure}[htp]
\centering
\includegraphics[scale=0.21]{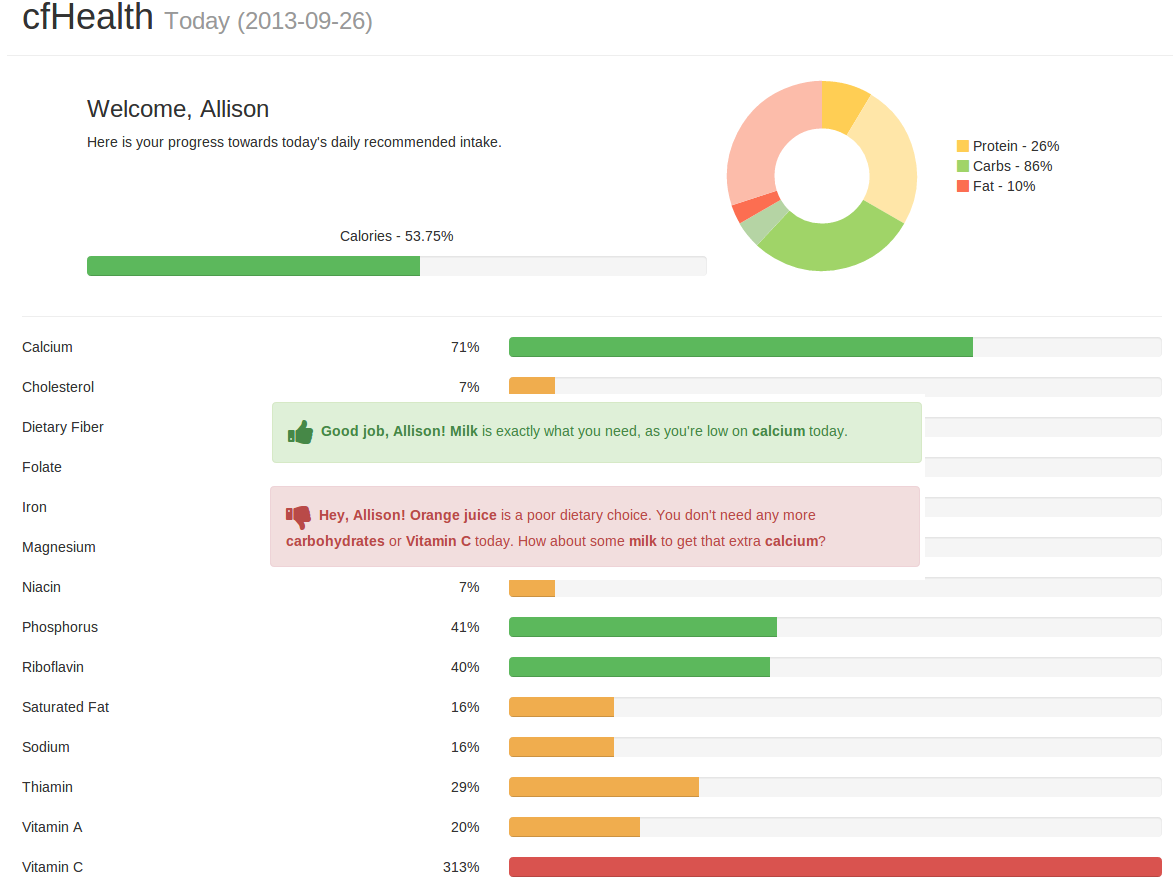}
\caption{Screenshot of cfHealth application.}\label{fig:health}
\end{figure}

The {\it CloudFridge Health} (cfHealth) application~\footnote{http://youtu.be/\_KnAJr\_7\_mY} monitors your food consumption and 
computes calorie intake and other health-related metrics. It shows the user their daily 
intake of various minerals and vitamins and compares them to the recommended daily intake. 
It then provides suggestions when the fridge door opens about what the user should eat 
to follow these recommendations. When the door closes there is either positive or negative
feedback based on what you chose (see Figure~\ref{fig:health}). For example, if a user has already consumed a high amount of 
carbohydrates but no calcium one day, the app will suggest the user to drink milk instead of 
coke and give an explanation why. This is done using a backend based on the WolframAlpha API, 
which is used to retrieve the nutritional information of the current content of the fridge and the
history of consumed food items. It is assumed that when an item is removed from the fridge one portion of 
the item is consumed. This application is also designed as a responsive mobile Web app.

The main benefit of this application is that you do not have to enter each item manually, which is common in similar health monitoring
applications. It could also be extended to offer custom diet profiles and combine currently stored items into recipes which 
would allow the user to follow their personal dietary goals more easily.

\subsection{Interaction-Design Lessons Learned}
Having access to the trace of items removed and the times when they were removed were crucial to implement our two sample 
applications, and we believe that there is a whole class of interesting applications that
can mine this trace both off-line and on-line. Our real-time event service allowed these applications
to trigger recommendations when they are needed without disturbing the user. 
In summary we have demonstrated that the following interaction scenarios can easily be
implemented on top of our testbed:
\begin{itemize}
\item search items by voice and direct your attention to their position,
\item alert you of expired items,
\item predict what you want to take out based on your task (e.g. ingredients when cooking a dish),
\item show you what is healthy for you to take out and why.
\end{itemize} 
What we note about these scenarios is that they deliver
recommendation at the right time at the right place, and they also all
adapt based on usage, and personal preferences. We argue that these features are inherent
in our system design and guide the application designs to meet the key criteria of
low-intrusion interactions.
We hope to learn more about similar interaction patterns when tracking actual usage
of item removals in user pilots with these applications in future work.

\section{Conclusion}\label{sec:conclusion}
We have demonstrated two use case implementations that support novel fridge interactions
based on a proposed testbed system. A lesson learned from these use cases is that casting the interaction as
an item-recommendation problem triggered when the user is about to take something out, can lead to
many new and interesting low-intrusion dialogs with the fridge.

We have also shown that our testbed provides good item and position recognition
accuracy (accuracy of 83-88\%) at a lower overhead (27\%) than a state-of-the-art barcode scanning
method.



\section*{Acknowledgments}
This work was supported in part by the IT R\&D program of MSIP/KEIT ({\it 10045459}).

\bibliographystyle{./IEEEtran}
\bibliography{./IEEEabrv,./related}

\end{document}